\documentclass[aps,prd,twocolumn,,nofootinbib,10pt,longbibliography]{revtex4-1}
\usepackage{blindtext}

\usepackage{amssymb,amsmath}
\usepackage{epsfig}
\usepackage[dvipsnames]{xcolor}
\usepackage[utf8]{inputenc}
\usepackage{stmaryrd}
\usepackage{mathrsfs}
\usepackage{mathalfa}
\usepackage{accents}
\usepackage[normalem]{ulem}

\usepackage{graphicx}
\graphicspath{ {c:/Documents/PhysicsNotes/GradResearch/Images/GR_Lens/} }

\newcommand{\A}{\mathcal{A}}

\newcommand{\pa}{\partial}

\newcommand{\be}{\begin{equation}}
\newcommand{\ee}{\end{equation}}
\newcommand{\bea}{\begin{eqnarray}}
\newcommand{\eea}{\end{eqnarray}}
\newcommand{\ba}{\begin{equation}\begin{aligned}}
\newcommand{\ea}{\end{aligned}\end{equation}}

\newcommand{\beg}{\begin{gather*}}
\newcommand{\eng}{\end{gather*}}
\newcommand{\hh}{,\hspace{0.5cm}}
\newcommand{\hhh}{,\hspace{0.2cm}}

\newcommand{\n}[1]{\label{#1}}

\newcommand{\CAL}{\mathcal}

\newcommand{\ts}[1]{{\boldsymbol{#1}}}

\def\XXint#1#2#3{{\setbox0=\hbox{$#1{#2#3}{\int}$ }
\vcenter{\hbox{$#2#3$ }}\kern-.6\wd0}}

\usepackage[makeroom]{cancel}
\usepackage[caption=false]{subfig}
\usepackage[colorlinks=true,
            citecolor=green,
            linkcolor=red,
            filecolor=cyan,
            urlcolor=magenta,
            backref=false]{hyperref}

\newcommand{\dw}{\dfrac{1}{\omega}}

\begin{document}

\title{Spinoptics in a curved spacetime}

\author{Valeri P. Frolov}%
\email[]{vfrolov@ualberta.ca}
\affiliation{Theoretical Physics Institute, Department of Physics,
University of Alberta,\\
Edmonton, Alberta, T6G 2E1, Canada
}


\begin{abstract}
In this paper we study propagation of the high frequency electromagnetic waves in a curved spacetime. We discuss a so call spinoptics approach which generalizes a well known geometric optics approximation and allows one to take into account  spin-dependent corrections. A new element proposed in this work is the use of an effective action which allows one to derive spinoptics equations. This method is simpler and more transparent than the earlier proposed methods. It is explicitly covariant and can be applied to an arbitrary spacetime background. We also demonstrate how the initial value problem for the high frequency electromagnetic wave can be solved in the developed spinoptics approximation.

\medskip

\hfill {\scriptsize Alberta Thy 4-24}
\end{abstract}

\maketitle

\section{Introduction}

Most information in the cosmology and astrophysics is obtained by observation of electromagnetic radiation.
In this paper we consider propagation of a high frequency electromagnetic wave in a curved spacetime and discuss how its polarization affects the wave propagation.
Before study of the problem in the curved background let us remind basic well known properties of   Maxwell field $F_{\mu\nu}$ in a flat spacetime. A general solution of source-free Maxwell equations can be presented as a superposition of the plane monochromatic waves, characterized by a 4D wave vector $\ts{k}$. For each  $\ts{k}$ there exist two linearly independent solutions with different polarizations. To identify the polarization state one can consider a following gedanken experiment. Consider an inertial observer with 4D velocity $\ts{U}$. Two vectors  $\ts{U}$ and $\ts{k}$ uniquely determine a 2D plane $\Pi$ orthogonal to both vectors. The electric field $E^{\mu}=F^{\mu\nu}U_{\nu}$ in the frame of the observer $\ts{U}$ lies in a plane $\Pi$. In a general case its "end point" moves along an ellipse. One can present such a wave as a superposition, of right- and left-circular polarized waves. In "quantum" language, these states are characterized by their helicity. It is easy to check that if the monochromatic plane wave is right- or left-polarized for the observer $\ts{U}$ then it has the same property for any other inertial observer. This  is in an agreement with Lorentz invariance of the helicity.

In many practically important situations the wavelength of the electromagnetic radiation is much shorter compared to the size of structures with which  it interacts. In such cases the geometric optics (high frequency) approximation
is very useful. Source-free Maxwell equations written for a vector potential $A_{\mu}$ in the Lorenz gauge in a curved spacetime have the form
\be \n{MAX}
J\equiv \Box A_{\mu}-R_{\mu}^{\ \nu}A_{\nu}=0\hh A^{\mu}_{\ ;\mu}=0\, .
\ee
In the geometric optics one searches for approximate solutions of the Maxwell equation with the following form of the 4D vector potential
\be \n{AAA}
A_{\mu}=\A_{\mu} \exp(i\omega S)\, .
\ee
Here the vector $\A_{\mu}$ and the scalar $S$ are slowly changing functions of coordinates, and $\omega$ is a constant which has dimension of the frequency.

Consider a local Lorentz frame and let $U^{\mu}$ be a unit timelike vector of the observer at rest in this frame. Then
\be
\omega_U=-\omega U^{\mu}S_{,\mu}
\ee
 is the frequency of the wave \eqref{AAA}  as measured by the observer $\ts{U}$. The frequencies  $\omega_U$  measured by different local Lorentz observers are different  at different locations. However, if one rescales $\omega\to C\omega$ all frequencies $\omega_U$ rescale similarly $\omega_U\to C \omega_U$. In the geometric optics one uses expansion of the amplitude in powers of $1/C$, or, what is the same, in powers of $1/\omega$. In particular, one writes
 \be\n{AA}
 \A_{\mu}=\A_{\mu}^{(0)}+\dfrac{1}{\omega} \A_{\mu}^{(1)}+\ldots \, ,
 \ee
 where $\ldots$ denote higher in $1/\omega$ terms. It should be emphasized that in this expansion the parameter $\omega$ simply keeps trace of the order of different terms under its rescaling.

 The Lorenz gauge condition in the leading order gives
 \be \n{LG}
 S^{,\mu}\A_{\mu}^{(0)}=0\, .
 \ee
 After substitution of \eqref{AAA} into \eqref{MAX} one gets
 \be
 J_{\mu}=-\omega^2\big[  J_{\mu}^{(0)}+\dfrac{1}{\omega} J_{\mu}^{(1)}+\dfrac{1}{\omega^2} J_{\mu}^{(2)}\big]=0\, .
 \ee
The condition $ J_{\mu}^{(0)}=0$ implies the eikonal equation
\be\n{SS1}
(\nabla S)^2=0\, .
\ee
The condition $ J_{\mu}^{(1)}=0$ implies the  equation
\be \n{SA}
S^{,\nu}\A_{\mu ;\nu}^{(0)}+\dfrac{1}{2} \A_{\mu}^{(0)}S^{,\nu}_{\ ;\nu}=0\, .
\ee
Relations \eqref{LG}, \eqref{SS1} and \eqref{SA} are basic equations of the  geometric optics,
while the third equation $ J_{\mu}^{(2)}=0$ can be used to get corrections to it.

In this approach, finding a solution of the Maxwell equations in the curved spacetime reduces to study of null rays propagation.This approach is nicely described in the MWT book \cite{MTW}, where many useful references can be found (for a more recent review see e.g. \cite{Dolan:2018nzc}). The main results can be formulated as follows. Let us write $\A_{\mu}^{(0)}$ for a linearly polarized wave in the form $\A_{\mu }^{(0)}=a b_{\mu}$, where $b^{\mu}$ is a real unit spacelike vector, then:
(1) $k_{\mu}=S_{,\mu}$ are null vectors. Their integral lines are geodesic null rays.
(2) The vector of polarization $b^{\mu}$ is orthogonal to $k^{\mu}$ and is parallel propagated along null rays.
(3) Scalar amplitude $a$ obeys the conservation law $\nabla_k(a^2{k}^{\mu})=0$.

To find a solution one uses the asymptotic expansion of the vector amplitude \eqref{AA}. However, in a practical application of the geometric optics there exists a following problem. Suppose one fixes some initial conditions for the wave and uses the geometric optics approximation to find a solution. Then even in a flat spacetime a domain of validity of this approximation depends on the choice of the initial conditions. For example, this approximation breaks down near caustics. In a curved spacetime the break down of this approximation may happen if in some domain  the curvature becomes large. It seems to be impossible to provide general covariant conditions  which define the domain where the geometric optics  properly approximates the exact solution of the Maxwell equation. Instead of this in MWT book \cite{MTW} some "physically reasonable" conditions are formulated when one can expect that the geometric optics does work well. Let us denote by $\lambdabar=1/\omega$ the typical reduced wavelength and  by $\ell$   the  typical length over which the amplitude, polarization and wavelength of the waves vary.
For example, for a beam of light $\ell$ is defined by its cross-section, while for the pulse of the radiation it depends on its duration. For waves $\ell$ can be connected with the radius of the curvature of the surface of the wave front.
Denote by  $R$ typical components of the Riemann tensor in some typical local Lorentz frame, and by
$\CAL{R}=|R|^{-1/2}$ the typical radius of the curvature of the spacetime through which the wave propagates. Then a condition of the validity of the geometric optics approximation can be formulated as
\be
\lambdabar\ll \mbox{min}(\ell,\CAL{R})\, .
\ee
Further details and explanations can be found in the book \cite{MTW}.

Long time ago using the geometric optics approximation, it was demonstrated that in the general case the polarization vector of the electromagnetic radiation propagating in a curved spacetime can rotate \cite{Skrotskii,Plebanski:1960}. This so called gravitational Faraday effect was discussed and studied for special cases in a number of publication \cite{PIRAN,Connors,Piran1985,Ishihara1988,Nouri_Zonoz:1999,Sereno:2004jx,Li_2022}.

There exists another interesting effect connected with the interaction of the spin of a photon with the curvature of the spacetime. Namely, for the light propagating on the curved spacetime, the photons with different helicity propagate along slightly different  paths. This effect is known as gravitational spin Hall effect (see e.g \cite{Oancea:2019pgm}).
To study this effect one should slightly modify ("upgrade") the standard geometric optics. Such a modification sometimes is called spinoptics.

The main idea of spinoptics approximation is the following. One starts with a standard ansatz similar to \eqref{AAA} for the vector potential with a complex null vector $\CAL{A}_{\mu}$ corresponding to the circular polarized state. The helicity dependent contribution appears only in the subleading terms of the $1/\omega$ expansion.
To take  into account the effect of the helicity on the light propagation,
one should include these helicity dependent terms in the eikonal equation. At short distances this modifies the rays trajectory only slightly. However, for a long distance such a spin effect is "enhanced". For two opposite circular polarizations the ray trajectories in an external gravitational field are slightly different. They are still null, but in a general case they are not geodesic. As a result a linearly polarized beam of radiation, which it is a superposition of opposite circular polarization states, splits and becomes a superposition of two spatially separated circular polarized beams.

Spinoptics in the special types of the gravitational fields (Kerr black holes and cosmology) was discussed in many publications (see e.g. \cite{Frolov:2011mh,Frolov:2012zn,Yoo:2012vv,Shoom:2020zhr,Oancea:2023ylb,Shoom:2024zep} and references therein). In papers \cite{Oancea:2019pgm,Oancea:2020khc,Frolov:2020uhn,Dahal:2022gop} a general formulation of the spinoptics equations in an arbitrary gravitational field was proposed. Spinoptics approximation for high frequency gravitational waves propagating in a curved spacetime background was developed in \cite{Yamamoto:2017gla,Andersson:2020gsj,Dahal:2021qel,Li:2022izh,Kubota:2023dlz}.
In the present paper we propose a different derivation of the spinoptics equations for the high frequency electromagnetic waves in a curved geometry. Instead of study the high-frequency expansion of the Maxwell equations, we use an approach which is similar to the one developed earlier in the paper \cite{Stachel:1977cm}.  Namely, by substituting into  the action for the electromagnetic field  the ansatz, similar to \eqref{AAA}, we derive an effective action.  We demonstrate, that such an effective action calculated in the high-frequency approximation,
which includes first order in $1/\omega$ frequency helicity sensitive terms,  allows one to obtain the spinoptics equations.

In this paper we use units in which $G=c=1$, and the signature conventions of the book \cite{MTW}.

\section{Effective action}

Let us formulate a problem. We study approximate high frequency solutions of the source-free Maxwell equations in a curved spacetime $\CAL{M}$ with metric $g_{\mu\nu}$
\be
F^{\mu\nu}_{\, \, ;\nu}=0\, .
\ee
For this purpose we proceed as follows.
Consider a complex Maxwell field $F_{\mu\nu}$ the action for which is of the form
\be \n{WWW}
W=\dfrac{1}{8}\int\CAL{F}^2\, \sqrt{-g}\, d^4x\hh
\CAL{F}^2= F_{\mu\nu}\bar{F}^{\mu\nu}
\, .
\ee
To develop the spinoptics approach we write a complex field potential  $A_{\mu}$  in the form
\be\n{AAAA}
A_{\mu}=a M_{\mu}e^{i\omega S}\, .
\ee
Here amplitude $a$ and phase $S$ are  real scalar functions. $M^{\mu}$ is a complex null vector, satisfying the following conditions
\be \n{MMM}
M_{\mu}{M}^{\mu}=\bar{M}_{\mu}\bar{M}^{\mu}=0\hhh M_{\mu}\bar{M}^{\mu}=1\hhh M^{\mu}S_{,\mu}=0\, .
\ee
As in the standard geometric optics a surface $S=$const defines a wavefront and the last condition in \eqref{MMM} means that the complex null vectors $\ts{M}$ and $\bar{\ts{M}}$ are tangent to these surfaces.
Simple calculations give the following expression for the complex field strength
\be\n{AFF}
\begin{split}
&F_{\mu\nu}=A_{\nu ;\mu}-A_{\mu ;\nu}=\Big[ i\omega a(S_{,\mu}M_{\nu}-S_{,\nu}M_{\mu})\\
& +[ (aM_{\nu})_{;\mu}-(aM_{\mu})_{;\nu}] \Big]\exp(i\omega S)\, .
\end{split}
\ee
Substituting this expression in $\CAL{F}^2$ and
keeping terms of the second and first order in  $\omega$ one obtains
\be \n{FFF}
\begin{split}
&\CAL{F}^2=4\omega^2 a^2\left[ \dfrac{1}{2} (\nabla S)^2-\dfrac{1}{\omega} B^{\mu}S_{,\mu}
\right]\, ,\\
&B_{\mu}=i\bar{M}^{\nu}M_{\nu ;\mu}\, .
\end{split}
\ee
It is easy to check that the vector $B^{\mu}$ is real.

In \eqref{FFF} we omit the term
\be
\dfrac{1}{\omega^2}[ (aM_{\nu})_{;\mu}-(aM_{\mu})_{;\nu}][ (a\bar{M}^{\nu})^{;\mu}-(a\bar{M}^{\mu})^{;\nu}]\, .
\ee
Similarly to the geometric optics approach, we assume that both the amplitude $a$ and the polarization vector $M_{\mu}$ change slowly and the characteristic scale of this change is of order of $\ell$. Under this assumption the omitted term is of the order of $(\lambdabar/\ell)^2$, and hence for small $\lambdabar$ (large frequency $\omega$) it is much smaller that the terms present in the expression for $\CAL{F}^2$ in \eqref{FFF}.

By substituting the expression \eqref{FFF} into the action \eqref{WWW} one obtains a functional which depends on three variables: the amplitude $a$, polarization vector $\bf{M}$ and the phase function $S$. We write $W=\dfrac{1}{2}\omega^2 I$, and use $I$ as the effective action for the spinoptics approximation. It differs from $W$ by a constant factor. This does not affect the equations of motion.

Since in the derivation of $\CAL{F}^2$ we used relations \eqref{MMM}, one should include the corresponding constraints in the  obtained effective action. Let us denote
\be
\begin{split}
&\Lambda=\dfrac{1}{2}\bar{\lambda}_1 M_{\mu}M^{\mu}+\dfrac{1}{2}\lambda_{1}\bar{M}_{\mu}\bar{M}^{\mu}+\lambda_{2} ({M}_{\mu}\bar{M}^{\mu}-1)
\\
&+\bar{\lambda}_{3} M^{\mu} S_{,\mu}+\lambda_{3}\bar{M}^{\mu}S_{,\mu}\, .
\end{split}
\ee
Here $\lambda_1$ and $\lambda_3$ are complex Lagrange multipliers, and $\lambda_2$ is a real one.
We include these constraints into the effective action $I$ and write in in the form
\be
I=\int d^4x \, \sqrt{-g}\, \Big[ a^2 [ \dfrac{1}{2} (\nabla S)^2-\dfrac{1}{\omega} B^{\mu}S_{,\mu}]+\Lambda\Big]\, .
\ee

\section{Equations}

Let us note that constraint $\Lambda$ does not depend on the amplitude $a$.
Since $a\ne 0$ the variation of $I$ with respect to $a$ gives the following equation
\be\n{HHH}
H=  \dfrac{1}{2}(\nabla S)^2-\dfrac{1}{\omega} B^{\mu}S_{,\mu}=0\, .
\ee
Let us denote
\be\n{JJ}
J^{\mu}=a^2(S^{,\mu}-\dw B^{\mu})+\bar{\lambda}_{3}M^{\mu}+\lambda_{3}\bar{M}^{\mu}\, .
\ee
Then the variation of the action $I$ over $S$ gives
\be\n{JJJ}
J^{\mu}_{\ ;\mu}=0\, .
\ee
Variation of $I$ with respect to $\bar{M}^{\mu}$ gives
\be\n{SMM}
-\dfrac{a^2 i}{\omega} S^{,\nu} M_{\mu ;\nu}+\lambda_{1} \bar{M}_{\mu}+\lambda_{2} M_{\mu}+\lambda_{3}S_{,\mu}=0\, .
\ee
By multiplying this relation by $M^{\mu}$ and using \eqref{MMM} one gets $\lambda_1=0$.

Let us note that the relations \eqref{MMM} are invariant under the rotation in the 2D plane spanned by the complex null vectors $\ts{M}$ and $\bar{\ts{M}}$
\be \n{psi}
M^{\mu}\to e^{i\psi} M^{\mu}\hh \bar{M}^{\mu}\to e^{i\psi} \bar{M}^{\mu}\, .
\ee
Under this transformation the vector $\ts{B}$ changes as follows $B_{\mu}\to B_{\mu}-\psi_{,\mu}$.
We use this freedom to put
\be\n{BBSS}
B^{\mu}S_{,\mu}=0\, .
\ee
Then by multiplying \eqref{SMM}  by $\bar{M}^{\mu}$ and using \eqref{MMM} and \eqref{BBSS} one gets $\lambda_2=0$.
Thus one can write \eqref{SMM} in the following form
\be \n{SS}
i\dfrac{a^2}{\omega}S^{,\nu} M_{\mu ;\nu}=\lambda_3S_{,\mu}\, .
\ee

Following the prescription of \cite{Stachel:1977cm} one should identify \eqref{HHH} with a Hamilton-Jacobi equation for a particle with 4-momentum $p_{\mu}=S_{,\mu}$
\be
H(p,x)=\dfrac{1}{2}g^{\mu\nu}p_{\mu}p_{\nu}-\dfrac{1}{\omega}B^{\mu}p_{\mu}\, .
\ee
This Hamiltonian determines a class of mechanical trajectories
associated with the high frequency electromagnetic waves.
The corresponding Hamiltonian equations are
\be \n{HAM}
\begin{split}
&\dfrac{d x^{\mu}}{ds}=\dfrac{\pa H}{\pa p_{\mu}}=p_{\mu}-\dfrac{1}{\omega}B_{\mu}\, ,\\
&\dfrac{d p_{\mu}}{ds}=-\dfrac{\pa H}{\pa x^{\mu}}\, .
\end{split}
\ee
Let us note, that equations \eqref{HAM} are similar to the equations of motion for a charges particle with the charge $1/\omega$ in the magnetic field with the potential $B_{\mu}$.

\section{Null ray congruence and complex null tetrad associated with it}

In order to formulate spinoptics equations we use a special complex null frame.
We denote by $l^{\mu}$ a tangent vector to the ray
\be\n{LB}
l^{\mu}=\dfrac{d x^{\mu}}{ds}=p^{\mu}-\dfrac{1}{\omega}B^{\mu}\, .
\ee
Then the second equation in \eqref{HAM} can be written as follows
\be \n{www}
w^{\mu}\equiv l^{\nu}l^{\mu}_{\, \, ;\nu} =\dfrac{1}{\omega} K^{\mu}_{\, \, \nu}l^{\nu}\hh K_{\mu\nu}=B_{\nu ;\mu}-B_{\mu ;\nu}\, .
\ee
This equation implies that
\be
l^{\mu}w_{\mu}=0\hh l^{\mu}(\ts{l}^2)_{,\mu}=0\, .
\ee
The latter equation shows that if $\ts{l}^2=0$ at some  moment of time,
then $\ts{l}^2$ vanishes along all the trajectory and the ray is null.

Let us consider a congruence of null rays and introduce a complex null tetrad $(l^{\mu},m^{\mu},\bar{m}^{\mu},n^{\mu})$ associated with it. These vectors are normalized as follows
\be
(\ts{m},\bar{\ts{m}})=-(\ts{l},\ts{n})=1\, ,
\ee
all other scalar products are zero.
For a given congruence $\ts{l}$ there exists the following freedom in the definition of such a null complex tetrad associated it and  preserving these normalization conditions
\begin{enumerate}
\item $\ts{l}\to \gamma \ts{l}\hh \ts{n}\to \gamma^{-1} \ts{n}$\, .
\item $\ts{m}\to e^{i\phi} \ts{m}$\hh $\bar{\ts{m}}\to e^{-i\phi} \bar{\ts{m}}$\, .
\item $\ts{l}\to  \ts{l}$,  $\ts{m}\to \ts{m}+\alpha \ts{l}$,  $\bar{\ts{m}}\to \bar{\ts{m}}+\bar{\alpha} \ts{l}$\, ,\\
 $\ts{n}\to \ts{n}+\bar{\alpha} \ts{m}+\alpha \bar{\ts{m}}  +\alpha\bar{\alpha}\ts{l}$\, .
\item $\ts{m}\to \bar{\ts{m}}$, $\bar{\ts{m}}\to \ts{m}$.
\end{enumerate}

One can use the normalization conditions to show that
\be
e^{\mu\nu\alpha\beta}l_{\mu}n_{\nu}m_{\alpha}\bar{m}_{\beta}=i \sigma\hh
\sigma=\pm 1\, ,
\ee
where $e^{\mu\nu\alpha\beta}$ is the Levi-Civita tensor.
We call a complex null tetrad right-handed if the helicity parameter $\sigma=1$, and left-handed if $\sigma=-1$. It is easy to see that the transformation 4 changes the helicity to the opposite one.

Let us note that under the transformation 1 the acceleration vector $\ts{w}$ changes as follows
\be\n{wl}
w^{\mu}\to \gamma^2 w^{\mu}+\gamma l^{\nu}\gamma_{,\nu} l^{\mu}\, .
\ee
Since $(\ts{w},\ts{l})=0$, \eqref{wl} implies that one can always find $\gamma$ such that the vector $\ts{w}$ takes the form
\be \n{kap}
w^{\mu}=\dfrac{1}{\omega}\kappa^{\mu}\hh \kappa^{\mu}=K^{\mu}_{\ \nu} l^{\nu}=\bar{\kappa} m^{\mu}+{\kappa} \bar{m}^{\mu}\, .
\ee
The complex function $\kappa$ in \eqref{kap} is of the order of 1.

In the limit $\omega\to\infty$  equation \eqref{www} reduces to a geodesic equation. We restrict the choice of the complex null tetrad by requiring that in the same limit its other vectors are also parallel propagated along the null ray. Thus one has
\be
Dm^{\mu}=\dfrac{1}{\omega}Z^{\mu}\hh
Dn^{\mu}=\dfrac{1}{\omega}N^{\mu}\, ,
\ee
where complex vector $Z^{\mu}$ and real vector $N^{\mu}$ are of the order of $\omega^0$.
The normalization conditions imply
\be\n{NNZZ}
\begin{split}
&(\ts{l},\ts{\kappa})=(\ts{n},\ts{N})=(\ts{m},\ts{Z})=(\bar{\ts{m}},\bar{\ts{Z}})=0\, ,\\
&(\ts{l},\ts{N})=- (\ts{n},\ts{\kappa})=0
\, , \ \  (\ts{l},\ts{Z})=- (\ts{m},\ts{\kappa})=-\kappa\,  ,\\
&(\ts{m},\ts{N})=-(\ts{n},\ts{Z})
\hh (\bar{\ts{m}},\ts{Z})=-(\ts{m},\bar{\ts{Z}})\, .
\end{split}
\ee

The vectors $N^{\mu}$ and $Z^{\mu}$  in the complex null frame have the form
\be\n{TNZ}
\begin{split}
&N^{\alpha}=N_l l^{\alpha}+ N_n n^{\alpha}+N_m m^{\alpha}+\bar{N}_{{m}} \bar{m}^{\alpha}\, ,\\
&Z^{\alpha}=Z_l l^{\alpha}+ Z_n n^{\alpha}+Z_m m^{\alpha}+Z_{\bar{m}} \bar{m}^{\alpha}\, .
\end{split}
\ee
Relations \eqref{NNZZ} impose restrictions on the coefficients in \eqref{TNZ}. Using them one gets
\be\n{TTNZ}
\begin{split}
&N^{\alpha}=\bar{Z}_l m^{\alpha}+Z_l \bar{m}^{\alpha}\, ,\\
&Z^{\alpha}=Z_l l^{\alpha}+ \kappa n^{\alpha}+iZ m^{\alpha}\, .
\end{split}
\ee
Here $Z$ is a real function and $Z_l$ is a complex one.
One still has freedom to use transformations 2 and 3, provided that
$\phi=\dw \phi_1$ and $\alpha=\dw \alpha_1$,
where $\phi_1$ and $\alpha_1$ are of the order of 1.
Under these transformations one has
\be\n{DNZ}
\begin{split}
&Dn^{\mu}\to Dn^{\mu}+\dw [ D\bar{\alpha}_1 m^{\mu}+D{\alpha}_1 \bar{m}^{\mu}]\, ,\\
&Dm^{\mu}\to Dm^{\mu}+\dw [ i D\phi_1 m^{\mu}+D\alpha_1 l^{\mu}]\, ,
\end{split}
\ee
By choosing $D\phi_1 =-Z$ and $D\alpha_1=-Z_l$
one can simplify expressions in \eqref{TTNZ} and present them in the form
\be
N^{\mu}=0\hh Z^{\mu}=\kappa n^{\mu}\, .
\ee
The above results means that for large $\omega$ one can choose the complex null tetrad $(\ts{l},\ts{n},\ts{m},\bar{\ts{m}})$ associated with the null congruence $\ts{l}$ so that the following set of equations is valid
\be \n{DDDD}
Dl^{\mu}=\dw (\bar{\kappa} m^{\mu}+{\kappa} \bar{m}^{\mu})\hhh Dn^{\mu}=0\hhh Dm^{\mu}=\dw \kappa n^{\mu}\, .
\ee

\section{Spinoptics equations}

Our goal now is write a set of the obtained earlier equations
\eqref{HAM}, \eqref{JJ}, \eqref{JJJ}, and  \eqref{SS} in terms of equations for the congruence of null rays and a complex null tetrads associated with it. We write the complex null vector $\ts{M}$ in the form
\be\n{MMMM}
M^{\alpha}=m^{\alpha}+\dfrac{1}{\omega} \mu^{\alpha}\, .
\ee
Normalization conditions \eqref{MMM} imply
\be\n{mmm}
(\ts{\mu},\ts{m})=0\hh (\bar{\ts{\mu}},\ts{m})+(\ts{\mu},\bar{\ts{m}})=0\, .
\ee
Using the definition of $\ts{B}$ one gets
\be
\begin{split}
&B^{\alpha}=b^{\alpha}+\dfrac{1}{\omega} \beta^{\alpha}\hh b_{\alpha}=i\bar{m}^{\nu} m_{\nu ;\alpha}\, ,\\
&\beta_{\alpha}=i(\bar{\mu}^{\nu} m_{\nu ;\alpha}+\bar{m}^{\nu} \mu_{\nu ;\alpha})
\, .
\end{split}
\ee

Since the difference between $B_{\mu}$ and $b_{\mu}$ is of the order of $O(\dw)$, one can simplify equations \eqref{LB} and \eqref{kap} by keeping in them terms up to the order $O(\dw)$.
One has \footnote{
The leading order contribution to the field strength \eqref{AFF} is $F^{(0)}_{\mu\nu}=-i\omega a(l_{\mu}m_{\nu}-l_{\nu}m_{\mu})\exp(i\omega S)$.
It is easy to check that
${}^{^*\!\!}F^{(0)}_{\mu\nu}\equiv \dfrac{1}{2}e_{\mu\nu\alpha\beta}F^{(0)\alpha\beta}=is F^{(0)}_{\mu\nu}$.
That is in the leading order $\ts{F}$ is self-dual (for $s=1$) or anti self-dual for $s=-1$.
}
\be\n{lllb}
\begin{split}
&l^{\mu}=p^{\mu}-\dfrac{1}{\omega}b^{\mu}\, ,\\
&\kappa^{\mu}=k^{\mu}_{\ \nu} l^{\nu}\hh
k_{\mu\nu}=b_{\nu ;\mu}-b_{\mu ;\nu}\, .
\end{split}
\ee
Condition \eqref{BBSS} implies that
\be \n{lob}
l^{\mu}b_{\mu}=O(\dw)\, .
\ee
The relations in the second line of \eqref{lllb} written in the explicit form give
\be \n{KKRR}
\begin{split}
&\kappa_{\mu}=i(m_{\alpha;\nu\mu}-m_{\alpha;\mu\nu})\bar{m}^{\alpha}l^{\nu}\\
&+i (\bar{m}_{\alpha ;\mu} Dm^{\alpha}-{m}_{\alpha ;\mu} D\bar{m}^{\alpha})
\end{split}
\ee
Equations \eqref{DDDD} imply that $D m^{\alpha}$ and $D \bar{m}^{\alpha}$
are of the order of $O(\dw)$, and hence in our approximation the terms in the second line of \eqref{KKRR} can be omitted. Using a formula for the commutator of the covariant derivatives one gets
\be
\kappa_{\mu}=iR_{\mu\nu\alpha\beta}m^{\alpha}\bar{m}^{\beta}l^{\nu}\, .
\ee

Relation $M^{\mu}S_{,\mu}=0$ in the leading order is satisfied, while in a subleading order it gives
\be \n{MULL}
l^{\alpha}\mu_{\alpha}=-m^{\alpha}b_{\alpha}\, .
\ee
Relations \eqref{mmm} and \eqref{MULL} do not specify uniquely $\ts{\mu}$. In particular, they remain valid under the transformation
\be \n{LLMU}
\mu^{\alpha}\to \mu^{\alpha}+\nu l^{\alpha}\, .
\ee

"Dynamical" equation \eqref{SS} gives
\be\n{DYN}
\dfrac{ia^2}{\omega^2}[D\mu^{\alpha}+\kappa n^{\alpha}+b^{\beta} m^{\alpha}_{\ ;\beta}]=\lambda_3(l^{\alpha}+\dw b^{\alpha})\, .
\ee
This relation shows that $\lambda_3$ is of the order $O(1/\omega^2)$.
This implies that the terms with the Lagrange multiplier $\lambda_3$ in equation \eqref{JJ} can be omitted and \eqref{JJJ} takes the form
\be \n{CONS}
(a^2 l^{\mu})_{;\mu}=0\, .
\ee
This relation has a simple interpretation: a number of photons in a beam
formed by null rays is conserved.

We denote $\lambda_3=(ia^2/2\omega^2)\lambda$.
Then in the leading order equation \eqref{DYN} takes the form
\be \n{DDMU}
D\mu^{\alpha}+\kappa n^{\alpha}+b^{\beta} m^{\alpha}_{\ ;\beta}=\lambda l^{\alpha}\, .
\ee
Using the ambiguity \eqref{LLMU} one can put $\lambda=0$.

\subsection{Helicity dependence}

In our derivation of the spinoptics equations we used a right-handed basis and adopted a definition
\eqref{MMMM} for the polarization vector $\ts{M}$. This choice corresponds to the positive helicity $\sigma=+1$ of the high-frequency waves. Let us note that the vector $b^{\mu}$ and other similar quantities  are helicity sensitive.
If one changes $\ts{m}\to \bar{\ts{m}}$ they are transformed as follows
\be
b_{\mu}\to - b_{\mu}\hhh k_{\mu\nu}\to - k_{\mu\nu}\hhh
\kappa_{\mu}\to -\kappa_{\mu}\hhh \kappa\to -\bar{\kappa}\, .
\ee
Using these properties one can write the spinoptics equations in the following general form which is valid for both signs of the helicity $\sigma=\pm 1$
\be\n{DYNTET}
\begin{split}
&Dl^{\mu}=i\dfrac{\sigma}{\omega} R^{\mu}_{\ \nu\alpha\beta}m^{\alpha}\bar{m}^{\beta}l^{\nu}\, ,\\
&Dn^{\mu}=0\hhh Dm^{\mu}=\dfrac{\sigma}{\omega} \kappa n^{\mu}\, ,\\
&D\mu^{\alpha}=-\sigma(\kappa n^{\alpha}+b^{\beta} m^{\alpha}_{\ ;\beta})\, ,\\
&(a^2 l^{\mu})_{;\mu}=0\, .
\end{split}
\ee

\section{Initial conditions}

We specify the initial conditions as follows. Consider a 3D spacelike surface $\Sigma$ imbedded in $\CAL{M}$. We denote by $x^{\mu}$ ($\mu=0,1,2,3)$) coordinates in the bulk spacetime $\CAL{M}$ and by $y^i$ ($i=1,2,3$) coordinates in $\Sigma$. Then $\Sigma$ is defined by the following equations $x^{\mu}=x^{\mu}(y^i)$.
The induced metric on $\Sigma$ is
\be
h_{ij}=\dfrac{\pa x^{\mu}}{\pa y^i} \dfrac{\pa x^{\nu}}{\pa y^j}g_{\mu\nu}\, .
\ee
We denote by $\ts{t}$ a future directed unit vector orthogonal to $\Sigma$. We denote also by $S_0$ the value of the phase function on $\Sigma$
\be
S_0(y^i)=S(x^{\mu}(y^i))\, .
\ee
Equation $S_0(y)=$const defines a position of the 2D surface of equal phase (wave surface) at the initial moment of time.

3D vector $S_{0 ,i}$ is orthogonal to the wave surface and is tangent to $\Sigma$. It determines a direction of propagation of the wave. Denote $h^{ij}S_{0 ,i}S_{0 ,j}=q^2$. Then a 3D unit vector in the direction of the wave propagation is $\hat{e}_i=S_{0 ,i}/q$. Its 4D form is defined by the conditions
\be
t^{\mu}e_{\mu}=0\hh \dfrac{\pa x^{\mu}}{\pa y^i}{e}_{\mu}=\hat{e}_i\, .
\ee
We define two future directed null vectors on $\Sigma$
\be
l_0^{\mu}=\dfrac{1}{\sqrt{2}} (t^{\mu}+e^{\mu})\hh
n_0^{\mu}=\dfrac{1}{\sqrt{2}} (t^{\mu}-e^{\mu})\, .
\ee
We denote by $\ts{m}_0$ the complex null vector at the initial time and  choose it to be tangent to $\Sigma$ and orthogonal to $l_0^{\mu}$ and $n_0^{\mu}$. We also specify the slowly changing function of the amplitude of the field on $\Sigma$: $a=a_0(y^i)$.

To obtain a solution of the Maxwell equations with given initial conditions in the spinoptics approximation one proceeds as follows.
One searches for a solution in the form
\be
A_{\mu}=a\Big(m_{\mu}+\dw \mu_{\mu}\Big) \exp(i\omega S)\, .
\ee
Using the initial conditions $(\ts{l}_0,\ts{n}_0,\ts{m}_0,\bar{\ts{m}}_0)$ on $\Sigma$ one should solve the "dynamical equations" in the first two lines of \eqref{DYNTET}
The null rays are defined by the equation $dx^{\mu}/ds=l^{\mu}$. We assume that the surface $\Sigma$ is chosen so that only one ray passes through each point in its vicinity. A null ray is defined by coordinates $y^i$ of the point on $\Sigma$ where the ray intersects it. We choose $s=0$ at $\Sigma$. Then $(y^i,s)$ specify a point on the ray.
Using relation $S_{,\mu}=l_{\mu}+(\sigma/\omega)b_{\mu}$ one can show in the spinoptics approximation the phase function $S$ remains constant on a given null ray up to the order of $O(1/\omega^2)$. Really, for two points 1 and 2 on the ray one has
\be
\Delta S=\int_1^2 dx^{\mu} S_{,\mu}=\int_1^2 ds\  l^{\mu}(l_{\mu}+\dfrac{\sigma}{\omega}b_{\mu})=O(1/\omega^2)\, .
\ee
To get this result we use relation \eqref{lob}. This means that the value of $S$ at point $x^{\mu}$ is equal to the value of the  phase function
$S_0$ a the point where a null ray passing through $x^{\mu}$ intersects $\Sigma$. The last equation in \eqref{DYNTET} determines the amplitude $a$ of the field.

Let us summarize. In this paper we studied propagation of the high-frequency electromagnetic waves in a curved space. We started with
with the ansatz \eqref{AAAA} which (locally) reproduces the complex vector potential for a monochromatic circularly polarized plane wave. After substitution of this ansatz into the action for the complex electromagnetic field and keeping the leading and first order sub-leading terms in its high frequency expansion, we derived an effective action. The variation of this effective action produces the  spinoptics equations. We demonstrated that the helicity dependent contribution enters in the sub-leading terms of these equations. We also demonstrated how using the spinoptics equations one can find a solution for an initial value problem for the high frequency electromagnetic ways. Let us emphasise that the effective action method developed in this paper is simpler and more transparent, then earlier  discussed derivation of the spinoptics equations based on the direct study of high frequency solutions of the Maxwell equations. The developed effective action approach allows a natural generalization for study spinoptics of gravitational waves propagating in a curved spacetime \cite{Frolov:2024qow}.

\section*{Acknowledgments}

The author thanks the Natural Sciences and Engineering Research Council of Canada and the Killam Trust for their financial support.



%

\end{document}